\documentclass[aps,prl,preprint,superscriptaddress]{revtex4-1}
\usepackage{graphicx}
\usepackage{graphics}
\usepackage{longtable}
\usepackage{float}
\usepackage{pst-circ}
\usepackage{float}
\usepackage{multirow}
\usepackage{array}
\begin{document}

%\preprint{}
\title{\bf{Exploring the origin of degenerate doublet bands in $^{106}$Ag}}


\author{N. Rather}
\affiliation{Saha Institute of Nuclear Physics, Kolkata-700064, INDIA}
\author{P. Datta}
\email[Corresponding author: ]{pdatta.ehp@gmail.com}
\affiliation{Ananda Mohan College, Kolkata-700009, INDIA}
\author{S. Chattopadhyay}
\affiliation{Saha Institute of Nuclear Physics, Kolkata-700064, INDIA}
\author{S. Roy}
\affiliation{Tata Institute of Fundamental Research, Mumbai-400005, INDIA}
\author{S. Rajbanshi}
\author{A. Gowsami}
\affiliation{Saha Institute of Nuclear Physics, Kolkata-700064, INDIA}
%\thanks{}
\author{G. H. Bhat}
\author{J. A. Sheikh}
\affiliation{Department of Physics, University of Kashmir, Srinagar-190006, INDIA}
\author{R. Palit}
\author{S. Pal}
\author{S. Saha}
\author{J. Sethi}
\author{S. Biswas}
\author{P. Singh}
\author{H. C. Jain}
\affiliation{Tata Institute of fundamental Research, Mumbai-400005, INDIA}
%\affiliation{ HENPP Division, Saha Institute of Nuclear Physics, Kolkata - 700064, INDIA}
%\affiliation{Tata Institute of Fundamental Research, Mumbai-400005, INDIA}
%\affiliation{Ananda Mohan College, 700009-Kolkata, INDIA}
\date{\today}

\begin{abstract}
The electromagnetic transition probabilities of the excited levels for the two nearly degenerate 
bands of $^{106}$Ag have been measured using the Doppler Shift Attenuation Method. A comparison with 
the calculated values using triaxial projected shell model approach indicates that these bands 
originate from two different quasi-particle configurations but constructed from the same mean-field deformation.
\end{abstract}

%\keywords{}

\maketitle
%\section{}
In the last decade, a number of nearly degenerate pairs of rotational bands with same parity 
have been reported in nuclei of mass A$\sim$130 \cite{1,2,3} and A$\sim$100 
\cite{4,5,6,7} regions. These bands are known to be strongly connected to each 
other and it has been proposed that a possible reason for the occurrence of these doublet bands 
is spontaneous breaking of chiral symmetry in triaxial nuclei due to the presence of three 
orthogonal angular momenta of the valence protons, valence neutrons and the core \cite{8,9}. 
However, for the two bands to be chiral partners, the near degeneracy in 
level energy and spin is a necessary but not a sufficient condition \cite{10}. In 
addition, these bands should exhibit nearly similar moment of inertia, quasi-particle alignment, signature staggering behaviour and, more importantly, the transition probabilities.

Indeed the nuclei of $^{134}$Pr \cite{1,2} and $^{104}$Rh \cite{4}, show the closest 
degeneracy in energy in the observed doublet bands over a wide angular momentum domain. However,
in both cases the quasi-particle alignment behaviour has been found to be different 
which indicates different shapes associated with the two bands. This has been supported by dissimilar behaviour of the measured B(E2) rates in the two bands of $^{134}$Pr \cite{11} which rules out the possibility of static chirality. The experimental transition rates could, however, be reproduced within 
an interacting boson-fermion-fermion model (IBFFM) framework by assuming a 
triaxial equilibrium deformation with fluctuation in shape around this value. Such a 
model would imply that the nuclear system will fluctuate between chiral and achiral 
configurations \cite{11}.

An alternate view on the origin of doublet bands has emerged based on the framework 
of the tilted axis cranking (TAC) model complemented by the random phase approximation (RPA) and its 
success in describing the experimental data in $^{135,136}$Nd \cite{12,13}. In this model, the doublet 
bands correspond to zero RPA phonon and the one RPA phonon configurations, respectively. The lowest 
RPA phonon energy accounts for the energy difference between the two bands near the band head and 
originates due to fluctuation in the orientation of shape perpendicular to the plane of neutron and 
proton angular momenta. Such chiral vibrations decreases with increasing angular momentum 
which results in the decrease of energy spacing between the doublet bands. This situation has been 
realized in $^{128}$Cs \cite{14} and $^{135}$Nd \cite{12} which may indicate the onset of chiral 
rotation at high spins. In both these nuclei, the transition rates for the doublet have been measured 
to be very similar and thus, are considered as the best candidates for chiral partner bands.

This picture opens up the possibility of a shape transformation in a $\gamma$-soft nucleus due to the chiral vibrations. In this case, the shape corresponding to the main band can be quite different from
its partner band (one phonon configuration). The experimental data on the doublet bands in 
$^{106}$Ag has been interpreted in this way where the possible deformation parameters for the main 
and the partner bands were found to be ($\beta$, $\gamma$) = (0.12, ${28}^0$) and (0.20, ${0}^0$), 
respectively, from total routhian surface (TRS) calculations \cite{5}. This picture gives an intuitive 
explanation for the existence of doublet bands with different moments of inertia and 
quasi-particle alignment behaviour. 

In a recent publication \cite{15}, Ma $et$ $al.$ have proposed that the doublet bands in 
$^{106}$Ag may originate due to two different quasi-particle structures, namely $\pi (g_{9/2})^1 
\otimes \nu (h_{11/2})^1$ for the main and $\pi (g_{9/2})^1 \otimes \nu (h_{11/2})^3$ for the 
partner band. However, the band crossing at $I$ = 18 $\hbar$ predicted by the Cranked Nilsson
Strutinsky (CNS) calculations, is substantially higher than the observed
crossing at 14 $\hbar$. In addition, a preliminary B(E2) transition rates measurement \cite{16} 
has been cited by Ma $et$ $al.$. These reported B(E2) values for partner band has large uncertainty of 
$\gtrsim$ 50$\%$. This measurement is consistent with both CNS prediction with 
fixed K value of 6 as well as with the previous interpretation \cite{5} of triaxial and axial 
prolate shapes of for the doublets. In addition, no B(M1) value has been reported which is essential to
determine the quasi-particle structure of the doublet bands. Thus, the previous investigations 
\cite{5,15} on the origin of the doublet bands of $^{106}$Ag have indicated two contrasting possibilities namely,
distinct shapes or distinct quasi-particle structures. 

In order to resolve this issue, we report the first accurate measurement of transition rates in the 
doublet bands of A$\sim$100 region in $^{106}$Ag. The observed transition rates of the two bands have 
been compared with the prediction of microscopic triaxial projected shell model (TPSM) \cite{17}. This 
model uses angular momentum projection technique to restore the rotational symmetry. Thus, it
provides the transition probabilities between states with well defined angular momentum. It is shown
that doublet bands originate from two different one-proton plus one-neutron quasi-particle
structures but arising from the same mean-field deformation.

The 68 MeV $^{14}$N beam from the 14-UD Pelletron at TIFR was used to populate the excited states of $^{106}$Ag through  $^{96}$Zr($^{14}$N, 4n) reaction. The 1 mg/cm$^2$ thick enriched $^{96}$Zr target had a $^{206}$Pb backing
of 9 mg/cm$^2$. The $\gamma$ rays were detected in Indian National Gamma Array (INGA) \cite{18}, which 
consisted of 20 Compton suppressed clover detectors. Two and higher fold coincidence data were recorded in fast
digital data acquisition system based on Pixie-16 modules of XIA LLC \cite{19}. The time stamped data was 
sorted in a $\gamma$-$\gamma$-$\gamma$ cube and three angle dependent $\gamma$-$\gamma$ matrices using 
the Multi-PArameter time stamped based Coincidence Search program (MARCOS), developed at TIFR.

The cube was analysed using the RADWARE program LEVIT8R \cite{20} to construct the level scheme of $^{106}$Ag and to determine the relative intensities of $\gamma$-rays which don't exhibit any lineshapes. The deduced level
structure for the doublet bands was found to be consistent with the reported level scheme \cite{5}. The partner
band was extended to ${I}^\pi$ = ${21}^-$ $\hbar$ through the addition of a 767 keV M1 transition. The crossover 
E2 transition of 1500 keV was also observed.

The angle dependent $\gamma$-$\gamma$ matrices were constructed by placing the gamma energies detected at specific angle (${40}^0$, ${90}^0$ and ${157}^0$) along one axis while the coincident $\gamma$ energy was placed on the other axis. The lineshapes were observed in $\gamma$-gated spectra for both E2 and M1 transitions above ${I}^\pi$ = ${12}^-$$\hbar$ and ${14}^-$$\hbar$ for the main and the partner bands, respectively. The relative intensities of these $\gamma$ rays were obtained from the gated spectrum at ${90}^0$ which were normalized with the intensities of the transitions of the same multi-polarity which have been estimated from the cube. The B(M1)/B(E2) values 
obtained from the relative intensities for the two bands matched with the previously reported values \cite{5} 
within $\pm$ 1 $\sigma$.

The lifetimes were measured at forward and backward angles by fitting the experimental lineshapes with the theoretical lineshapes derived from the code LINESHAPE by Wells and Johnson \cite{21}. The codes were used to generate the velocity profile of the recoiling nuclei at ${40}^0$ and ${157}^0$ with respect to the beam direction using the Monte Carlo technique with a time step of 0.001 ps for 5000 histories.
Northcliff and Shilling  \cite{22} stopping powers formula with shell correction was used for calculating the energy loss of the recoiling  $^{106}$Ag nuclei in $^{206}$Pb backing. The detailed procedure for lineshape fitting is described in ref. \cite{23,24}.

For the main band, the effective lifetimes for the ${18}^-$ and ${19}^-$ levels were found by fitting the lineshapes for 1309 (${19}^-$$\rightarrow$${17}^-$) and 1300 (${18}^-$$\rightarrow$${16}^-$) keV $\gamma$-transitions by assuming 100\% side-feed. The top feed lifetime for ${17}^-$ level was assumed to be the intensity weighted average of the lifetimes for ${18}^-$ and ${19}^-$ levels since this level is fed by both 674 (${18}^-$$\rightarrow$${17}^-$) and 1309 keV $\gamma$ rays. The side feeding intensity at this level was fixed to reproduce the observed intensity pattern at ${90}^0$ with respect to the beam direction. In this way each lower level was added one by one and fitted until all the observed lineshapes for 1206 (${16}^-$$\rightarrow$${14}^-$), 1146 (${15}^-$$\rightarrow$${13}^-$),  1042 (${14}^-$$\rightarrow$${12}^-$) and 979  (${13}^-$$\rightarrow$${11}^-$) $\gamma$ rays were included into a global fit where only the in-band and  feeding lifetimes were allowed to vary.  This procedure was repeated for ${157}^0$. The lifetime for the ${12}^-$ level was measured by fitting the lineshape of 833 (${12}^-$$\rightarrow$${10}^-$) keV $\gamma$ ray extracted from the top gate of 490 keV. This was done to avoid the large feed to this level from other non-yrast states. In this case, the observed lineshape was fitted by taking into account the complete top cascade but no side feeding at  ${12}^-$ level was considered. The uncertainties in the measured lifetimes were derived from the behaviour of ${\chi}^2$ fit in the vicinity of the minimum.

All these levels decay by M1 transitions and their lineshapes were also fitted following the same prescription
except for 490 (${13}^-$$\rightarrow$${12}^-$) and 489 (${12}^-$$\rightarrow$${11}^-$) keV $\gamma$ rays since
their lineshapes overlapped. Thus, the final values for the level lifetimes were obtained by taking the 
averages  from the fits at the two angles and for the two de-exciting transitions. The corresponding 
uncertainty in a level lifetime has been calculated as the average of the uncertainties for the independent
lifetime measurements for that level added in quadrature.

The same method of analysis was followed for the levels of the partner band. The ${12}^-$ level lifetime was
extracted from 326 keV M1 transition since the corresponding lineshape of cross-over E2 transition of 597 
keV could be contaminated due to the presence of Ge(n, n$^\prime \gamma$) reaction.

The examples of the lineshape fits in $^{106}$Ag are shown in Fig. 1. The B(M1) and B(E2) transition rates 
have been extracted from the measured level lifetimes and are tabulated in Table I. The error bars on the values
include the uncertainties on lifetime and intensity measurements added in quadrature. The measured rates are 
also plotted as a function of angular momentum in Fig. 2.

It is observed from Fig. 2 (a) that within the experimental uncertainty, the B(E2) rates for the two bands are
essentially same. This observation is in contradiction to the preliminary results 
quoted in Ref. \cite{15} which reported substantially larger B(E2) rates in the partner band. The present experimental data do not support the existing two 
explanations \cite{5,15} for the origin of the doublet bands in $^{106}$Ag since, in both the 
cases the B(E2) rates in the partner band are expected to be two times stronger than that in the main band. 

The B(M1) behaviour of the two bands are different as observed from Fig. 2 (b). The values for the partner band
exhibits staggering whose phase inverts around ${I}^\pi$ =  ${17}^-\hbar$. This staggering is absent for the main
band. It may also be noted that the B(M1) value shows a dip and a peak at ${I}^\pi$ =  ${17}^- \hbar$ for the 
main and partner band, respectively. Such differences in B(M1) values would indicate two different 
quasi-particle structures for the two bands.

We have compared the experimental transition rates of the two bands with the predictions of
TPSM. It is to be noted that this model has been successful in describing the chiral band structure and 
transition rates in $^{128}$Cs \cite{25} and the level structure and branching ratios of the doublet bands in 
$^{108}$Ag \cite{7}. For the present case of $^{106}$Ag, the deformation parameters for the triaxial
Nilsson potential used are $\epsilon$ = 0.15 and $\gamma$ = ${30}^0$. This parameter set is 
consistent with the previous TRS calculation \cite{5} and observed systematic of Ag-isotopes \cite{26}. 
For $^{106}$Ag, the nearly degenerate lowest two projected bands are found to have K = 4 and 2 for 
$\pi (g_{9/2})^1 \otimes \nu (h_{11/2})^1$ configuration. These bands are obtained from
angular-momentum projection of the triaxial deformed Nilsson intrinsic states. The projected states are, 
therefore, constructed from the same mean-field deformation.
The K = 2 band is $\sim$ 0.5 MeV higher at low spins (4$\hbar$-12$\hbar$) but becomes nearly 
degenerate with K = 4 bands around $I$ = ${14}\hbar$ which interestingly is the observed band crossing
spin for the doublet. These bands originate from different quasi-particle states and the Nilsson 
triaxial quasi-particle energies for K = 4 and K = 2 bands were  found to be 1.76  and 2.12 MeV, respectively.

In the second stage of the TPSM study, the projected bands are employed to diagonalise the
shell model Hamiltonian consisting of pairing and quadrupole-quadrupole interaction terms \cite{17}.
The energies for the doublet bands following diagonalisation, are shown in Fig. 3(a). It is quite
evident from the figure that calculated values are in good agreement with the 
experimental data and the band crossing spin is also been reproduced. It is to be noted that the CNS
calculations \cite{15} which assumes a four quasi-particle structure for the partner band does
not predict this crossing correctly.

The calculated amplitude probabilities of the wavefunctions ($a^2_{ik}$) for the main and the 
partner bands are plotted in Fig. 3(b) and (c)
as a function of spin. It is observed from the figure that at low spins, the main band predominantly
originates from K = 4 (1.76 MeV) configuration while the partner band originates from 
K = 2 (2.12 MeV) configuration. At higher spins ($I\geq $ 15$\hbar$), the partner band is pushed lower 
than the main band with K=4 configuration due to the large overlap of the degenerate K = 2 and 4 bands 
with same quasi-particle structure.

It is known that transition probabilities are
very sensitive to the wavefunctions and in order to confirm the above predicted two quasi-particle
structures for the doublet bands, it is quite important to compare the TPSM calculated
transition probabilities with the observed values. The comparison is shown in Fig. 2 and
it is evident that experimental B(E2) transition rates for the doublet bands are in 
good agreement with the predicted values. The observed drop at $I$ = ${16}$ $\hbar$ is due to the 
crossing between the doublet bands which leads to the loss of collectivity. 
The TPSM calculations also reproduce the observed variations of B(M1) rates as function of spin
for the doublet bands which are sensitive to their intrinsic quasi-particle configurations.

In summary, the level lifetimes of doublet bands have been reported for the first time in A$\sim$100 region for $^{106}$Ag. The previous works proposed that these bands correspond to two different shapes (triaxial and axial prolate) which originate due to $\gamma$-softness \cite{5} or two different
quasi-particle structures, $\pi (g_{9/2})^1 \otimes \nu (h_{11/2})^1$ and 
$\pi (g_{9/2})^1 \otimes \nu (h_{11/2})^3$ \cite{15}. However, the observation of very similar B(E2)
rates in the two bands is inconsistent with these interpretations. The good agreement between the
measured and the calculated transition rates obtained using TPSM, indicate that the doublet bands
correspond to same mean-field triaxial shape but originate from different quasi-particle configurations arising from $\pi (g_{9/2})^1 \otimes \nu (h_{11/2})^1$. These calculations also reproduce the crossing
between the main and partner bands at $I$ = 14$\hbar$.

Thus, the present measurement of transition probabilities in doublet bands of $^{106}$Ag
opens up the possibility that these bands in A$\sim$100 region and probably in other mass
regions may originate from different quasi-particle configurations. The lifetime measurements 
of the other observed doublet bands and a detailed study using the TPSM is highly desirable to 
explore this possibility.

Authors would like to thank the technical staff of TIFR-BARC pelletron facility for its smooth operation throughout the experiment. The help and cooperation of members of the INGA collaboration for setting up the array are acknowledged. This work was partially funded by the Department of Science and Technology, Government of 
India (No. IR/S2/PF-03/2003-III). N. Z. and P. D. (grant no. PSW-26/11-12) would also like to thank UGC for research support.

\newpage
\begin{figure}[h]
\includegraphics[scale=0.6, angle=0]{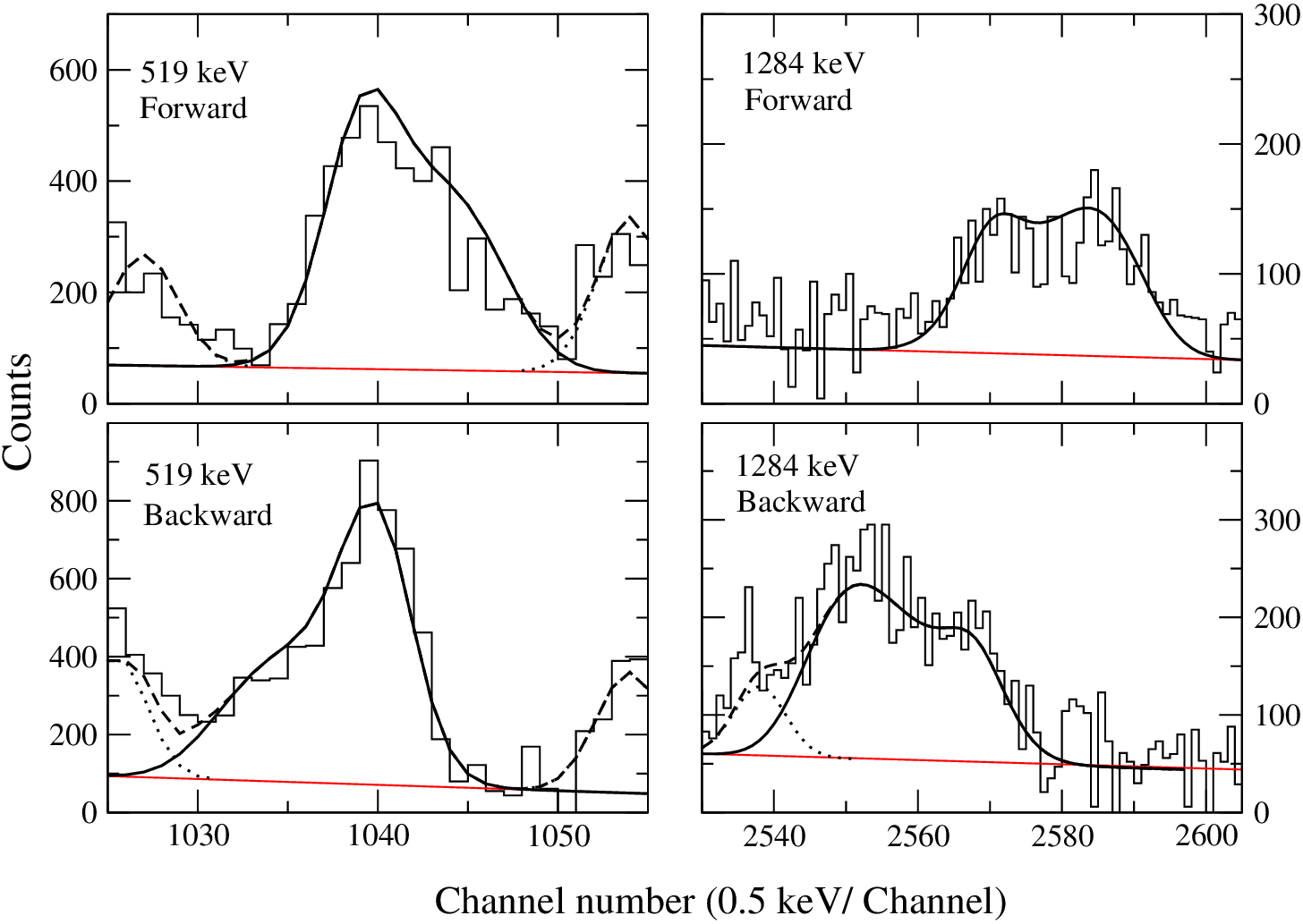}
\caption{Examples of the lineshape fits for 1284 (${19}^-\rightarrow{17}^-$) keV and 
519 (${17}^-\rightarrow{16}^-$) keV transitions  at ${40}^0$ and ${157}^0$ with respect 
to the beam direction. The Doppler broadened lineshapes are drawn in solid lines while the contaminant peaks are shown in dotted lines. The result of the fit to the experimental data is shown in dashed lines.}
\end{figure}

\newpage
\begin{figure}[h]
\centering
\includegraphics[scale=0.7, angle=0]{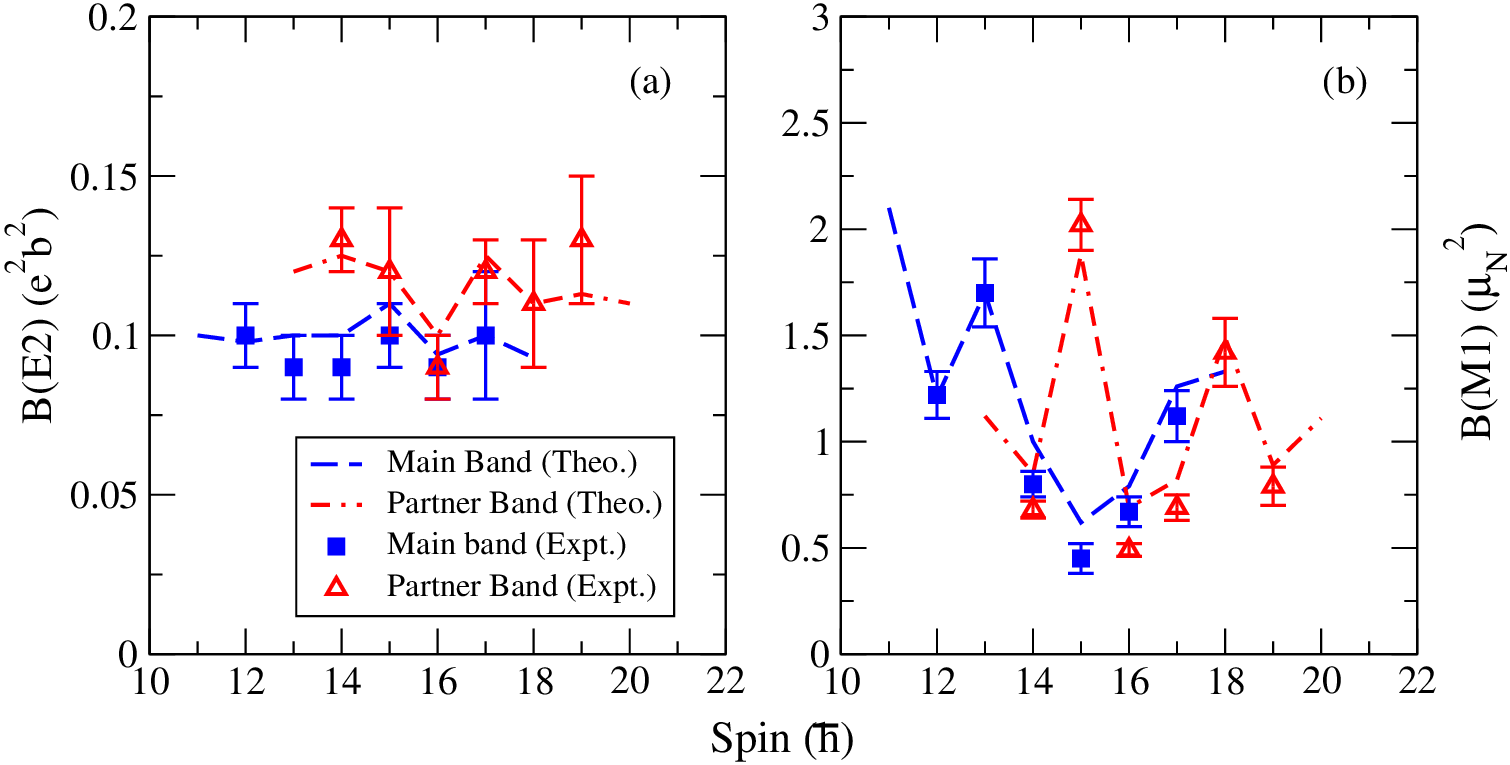}
\caption{Calculated and measured B(E2) (a) and B(M1) (b) rates for the doublet bands of $^{106}$Ag. Error bars on measured values of a given level include errors in intensity and lifetime of the level added in quadrature.} 
\end{figure}

\newpage
\begin{figure}[h]
\centering
\includegraphics[scale=0.8, angle=0]{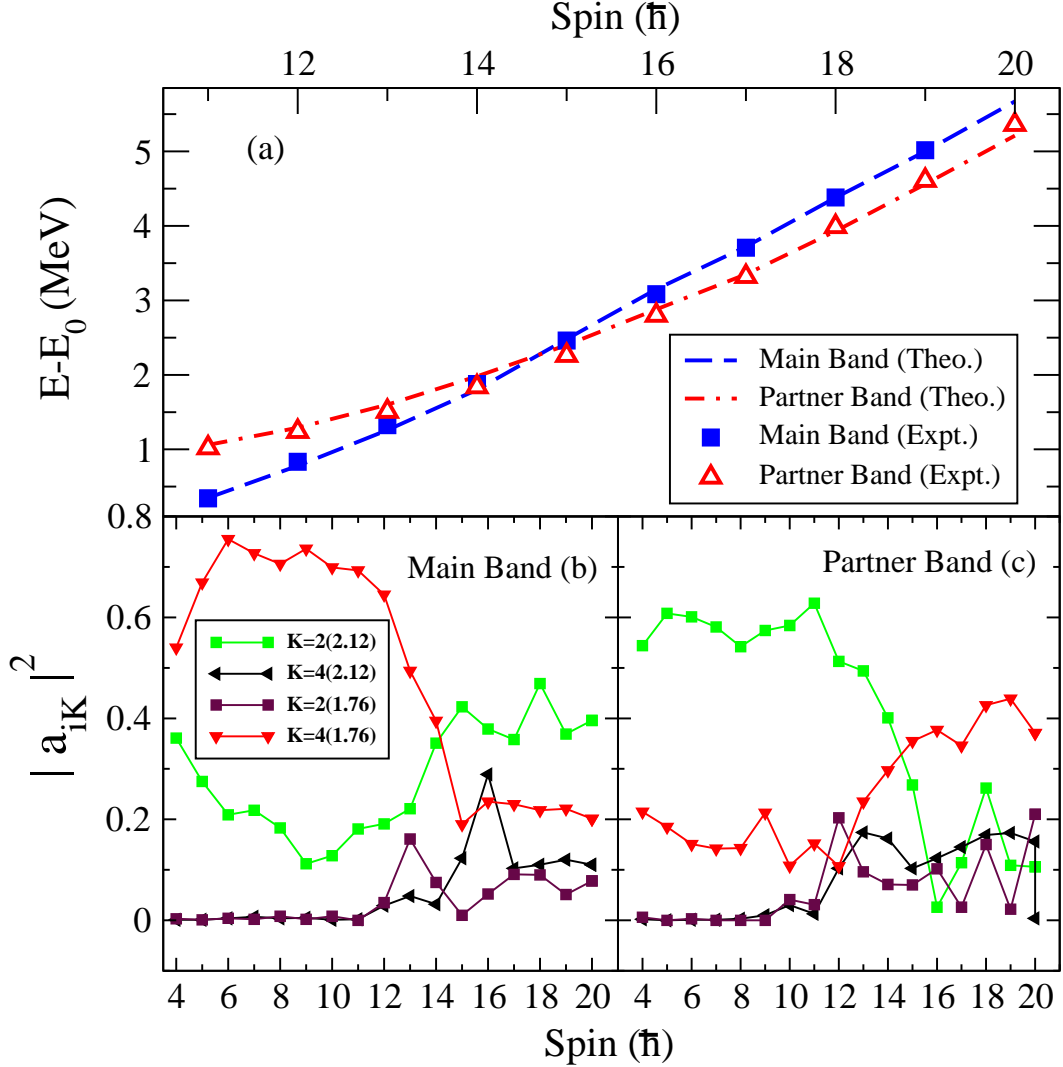}
\caption{(a) Comparison of the measured energy levels of negative parity doublet bands of $^{106}$Ag with those from TPSM calculation. The energies are relative to the band head $E_0$ taken to be the energy of ${I}^\pi$ = ${10}^-$$\hbar$ level of the main band. The calculated wavefunction amplitude probabilities from TPSM as a function of spin for the main and the partner bands are shown in 
(b) and (c), respectively. The numbers quoted in the parenthesis in the box of Fig. 3(b) are the quasi-particle
energies in MeV.} 
\end{figure}

\newpage
\begin{table}[h]
\centering
\begin{tabular}{cccc} \hline

Spin $I$[$\hbar$] &  Lifetime [ps]&  B(M1) [$({\mu_N}^2)$] &  B(E2) [${(eb)}^2$]\\ \hline
 & &Main Band \\ 
 ${12}^-$ &  0.32 (2) &  1.22 (1) &  0.10 (1) \\ 
 ${13}^-$ &  0.21 (2) &  1.70 (2) &  0.10 (1) \\  
 ${14}^-$ &  0.25 (2) &  0.80 (6) &  0.10 (1) \\ 
 ${15}^-$ &  0.23 (2) &  0.45 (7) &  0.10 (1) \\ 
 ${16}^-$ &  0.17 (1) &  0.67 (7) &  0.09 (1) \\ 
 ${17}^-$ &  0.12 (1) &  1.12 (12)&  0.10 (2) \\ %\hline
%%%end of yrast band%%
 & & Partner Band  \\ 
 ${14}^-$ &  1.77 (10)&  0.68 (4) &  0.13 (1) \\ 
 ${15}^-$ &  0.31 (1) &  2.02 (12)&  0.12 (2) \\
 ${16}^-$ &  0.42 (2) &  0.49 (3) &  0.09 (1) \\ 
 ${17}^-$ &  0.26 (1) &  0.69 (6) &  0.12 (1) \\ 
 ${18}^-$ &  0.09 (1) &  1.42 (16)&  0.1(2) \\
 ${19}^-$ &  0.11 (1) &  0.79 (9) &  0.13(2)\\ \hline
\end{tabular}
\caption{The measured lifetimes and the corresponding B(M1) and B(E2) values for the doublet bands of $^{106}$Ag.}
\label{table:table}
\end{table}

\end{document}